\documentstyle[12pt]{article}
\begin{document}

\rightline{CU-TP-778}
\rightline{hep-th/9608185}
\vskip 2cm
\centerline{\large\bf The Moduli Space of BPS
Monopoles\footnote{To be appeared in the proceedings of the Quark-96
workshop at Yaroslavl, Russia}}

\vskip 1.5cm

\centerline{\it
Kimyeong Lee\footnote{electronic mail: klee@phys.columbia.edu} }

\vskip 3mm
\centerline{Physics Department, Columbia University, New York, NY
10027, USA}
\vskip 1.5cm
\centerline{\bf ABSTRACT}
\vskip 5mm
\begin{quote}
We  review a progress in our understanding of the moduli
space for an arbitrary number of BPS monopoles in a gauge theory with
a group $G$ of rank $r$ that is maximally broken to $U(1)^r$. 
The derivation of the moduli space metric has been obtained from
studying the low energy dynamics of well-separated dyons. 
\end{quote}

\newpage
\setcounter{footnote}{0}
\section{Introduction}


Among the many remarkable features of the
Bogomol'nyi-Prasad-Sommerfield (BPS) limit \cite{bps} is the existence
of families of degenerate static multimonopole solutions.  For any
given topological charge, the space of such solutions, with gauge
equivalent configurations identified, forms a finite dimensional
moduli space.  A metric for this space is defined in a natural way by
the kinetic energy terms of the Yang-Mills-Higgs Lagrangian.  As was
first pointed out by Manton \cite{manton}, a knowledge of this metric
is sufficient for determining the low energy dynamics of a set of
monopoles and dyons.

The electromagnetic duality proposed by Montonen and Olive
\cite{montonen} in Yang-Mills Higgs systems can be realized in the
$N=4$ supersymmetric Yang-Mills-Higgs systems, which have the BPS
magnetic monopoles as solitons.  The duality implies that when the
coupling constant is large there is a dual theory where the role of
monopoles and elementary charged quanta is exchanged.  This means an
exact match between the spectrum of charged particles and that of
magnetic monopoles.  This in turn implies the existence of various
threshold bound states of fundamental magnetic monopoles. To see such
bound states of zero bound energy, one needs to understand the low
energy dynamics of BPS monopoles, which can be approached by the
moduli space approximation. (For the recent reviews of the
electromagnetic duality, see Ref.\cite{olive}.)

Thus it would be  interesting to understand the moduli space of the BPS
magnetic monopoles better. For the case of an $SU(2)$ gauge symmetry
spontaneously broken to $U(1)$, the moduli space ${\cal M}$ of
solutions carrying $n$ units of magnetic charge has $4n$ dimensions.
This naturally suggests that these solutions should be interpreted as
configurations of $n$ monopoles, each of which is specified by three
position coordinates and a $U(1)$ phase angle.  (Recall that time
variation of this phase angle gives an electrically charged dyon.)
The metric for the two-monopole moduli space was determined by Atiyah
and Hitchin \cite{atiyah}. By using this metric, the threshold dyonic
bound states of two identical monopoles have been found by Sen and
others\cite{sen}.  For three or more monopoles, the moduli space metric is
still unknown.  An asymptotic form, valid in the regions of ${\cal M}$
corresponding to widely separated monopoles, has been found by Gibbons
and Manton \cite{gibbons1}, but this develops singularities if any of
the intermonopole distances becomes too small, and hence cannot be
exact. The dyonic bound states of $n$ identical monopoles have
been discussed also \cite{porrati}.

In this talk, which is based on our work \cite{klee1,klee2}, we
consider in detail the moduli space of the BPS monopoles in the theory
with an arbitrary gauge group $G$ of rank $r\ge 2$ in the maximal
symmetry breaking (MSB) case where the unbroken gauge symmetry will be
a purely abelian subgroup $U(1)^r$.  Hopefully this talk is
complementary to Erick Weinberg's talk where he draw more general
pictures of the moduli space, duality and unbroken gauge symmetry.  In
this MSB case, there are $r$ independent topological charges $n_a$
and, correspondingly, $r$ {\it fundamental monopoles}, each of which
carries a single unit of one of these charges \cite{weinberg1}.  The
moduli spaces corresponding to any combination of $n$ fundamental
monopoles are $4n$-dimensional, just as in $SU(2)$.  The moduli space
metric for all two-monopole solutions has been found\cite{klee1},
which generalized the $SU(3)$ result outlined by us in \cite{klee2}
and also found by Gauntlett and Lowe\cite{gaunt} and by Connell
\cite{connell}.  In addition, the asymptotic form of the metric for
any number of monopoles is also found \cite{klee1}.  In the case where
all monopoles are distinct from each other this asymptotic form has
been argued to be the exact metric over the entire moduli
space\cite{klee1,murray,chalmers}

It might seem odd to try to obtain the moduli space metric for three
or more monopoles with a larger gauge group when this cannot even be
done for $SU(2)$.  The reason that it might actually be easier to work
with a larger group is that in a number of such cases the moduli space
possesses greater symmetry.  Specifically, consider a multimonopole
solution with $r$ fundamental monopoles, each corresponding to a
different topological charge.  The corresponding $U(1)$ factors of
the unbroken gauge group act nontrivially on this solution, and this
action generates an isometry on the moduli space.  Furthermore, for
any such collection of topological charges there is a solution that,
although it is composite, is spherically symmetric.  On the moduli
space, such a solution corresponds to a fixed point under the
isometries that correspond to overall rotation of the monopole
configuration.  In $SU(2)$ there are no spherically symmetric
solutions with multiple magnetic charge \cite{guth}, and hence no such
fixed points.

However generally the gauge symmetry might be partially broken to
$K\times U(1)^{r-k}$ with unbroken nonabelian group $K$ is a
semisimple group of rank $k<r$.  In this case there are many oddities
and one can define the moduli space of massive and massless monopoles
consistently only when total magnetic charge is purely abelian. The
the detail aspects of this case and its implications in the
generalization of the electromagnetic duality are discussed in Erick's
talk and Ref.\cite{klee3}. 

The plan of the talk is as follows. First, we recall some properties
of BPS monopoles in general gauge group $G$ in Sec.~2. Here we also
consider some characteristics of the moduli space metric.  We then
expose our detail analysis of the moduli spaces in Sec.~3.  Here,
following the approach that Gibbons and Manton \cite{gibbons1} used
for $SU(2)$, we use our knowledge of the interaction between widely
separated dyons to infer the asymptotic form of the moduli space
metric, which turns out to be smooth everywhere when the monopoles are
distinct.  Section 4 contains some concluding remarks.

\section{BPS Monopoles}

We begin by recalling some properties of the BPS solutions \cite{bps}
in an $SU(2)$ theory spontaneously broken to $U(1)$.  We fix the
normalization of the electric and magnetic charges $g$ and $q$ by
writing the large distance behavior of the electromagnetic field
strength (in radial gauge) as
\begin{equation} 
 B_i^a = \frac{ {\hat r}_i{\hat r}_a g}{ 4\pi r^2},\quad
 E_i^a = \frac{ {\hat r}_i {\hat r}_a q}{ 4\pi r^2} .
\end{equation}
The dyon solution carrying one unit of magnetic charge (i.e.,
$g=4\pi/e$, with $e$ being the gauge coupling) may be written as
\begin{eqnarray} 
\Phi^a({\bf r}) &=& {\hat r}^a K(r; v),\nonumber \\
A_i^a ({\bf r}) &=& \epsilon_{iak} {\hat r}^k \left(A(r; v) -
   \frac{1}{ er} \right),    \nonumber \\
A_0^a ({\bf r}) &=& \frac{q}{ \sqrt{g^2 +q^2}} \Phi^a,
\label{dyonsolution}
\end{eqnarray}
where $v$ is the asymptotic magnitude of the Higgs field and 
\begin{eqnarray} 
K(r;v) &=&  v \coth(evr \eta)  - \frac{1}{ er \eta},\nonumber \\
A(r; v) &=& \frac{ v \eta}{ \sinh(evr\eta) }  ,
\end{eqnarray}
with $ \eta = g/ \sqrt{g^2 +q^2} $. 
The mass of this dyon is $\tilde{m} = v  \sqrt{g^2 +q^2}$. after
quantization, the electric charge $q$ should be an integer multiplet
of $e$.

For an arbitrary gauge group $G$ of the rank $r$, its generators can be
chosen to be the $r$ Cartan subalgebra generators $H_i$ and the
lowering and raising operators, $E_{\mbox{\boldmath $\alpha$}}$ one for
each root $\alpha$.
They are  normalized so that
\begin{equation} 
{\rm tr}\, H_i H_j =\delta_{ij}, \,\,\,\, {\rm tr }\, E_{-\mbox{\boldmath
$\alpha$}}\, E_{\mbox{\boldmath $\beta$}} = \delta_{\mbox{\boldmath
$\alpha$}\mbox{\boldmath $\beta$}}
\end{equation}  
in a, say, adjoint representation.  The roots ${\mbox{\boldmath
$\alpha$}}$ may be viewed as vectors forming a lattice in a
$k$-dimensional Euclidean space.  It is always possible to choose a
basis of $k$ simple roots for this lattice in such a way that all
other roots are linear combinations of the simple roots with integer
coefficients all of the same sign; a root is called positive or
negative according to this sign.  A set of simple roots of particular
importance is defined as follows.  Let $\Phi_0$ be the asymptotic
value of the Higgs field in some fixed direction (say, the positive
$z$-axis).  We may choose $\Phi_0$ to lie in the Cartan subalgebra and
then define a vector $\bf h$ by
\begin{equation}     
       \Phi_0 =  {\bf h}\cdot  {\bf H} .
\end{equation}
Here we are first concerned with the case of maximal symmetry
breaking, with the gauge group spontaneously broken to $U(1)^k$.  This
is achieved if and only if $\bf h$ has a nonzero inner product with
all of the roots.  When this is so, there is a unique set of simple
roots ${{\mbox{\boldmath $\beta$}}}_a$ that satisfy the requirement
that ${\bf h} \cdot {\mbox{\boldmath $\beta$}}_a $ be positive for all
$a$; we shall use this basis for the remainder of this talk.

Asymptotically, the magnetic field must commute with the Higgs field.
Hence, in the direction chosen to define $\Phi_0$, it must be of the
form 
\begin{equation} 
{\bf B} =  \frac{{\hat r}}{ 4\pi r^2 } {\bf g}\cdot {\bf H } .
\label{magf}
\end{equation}
Topological arguments lead to the quantization condition \cite{topology}
\begin{equation}    {\bf g} = g \sum_{a=1}^k  n_a 
{{\mbox{\boldmath $\beta$}}}_a^* ,
\label{mcharge}
\end{equation}
where again $g=4\pi/e$ and 
\begin{equation}      
{{\mbox{\boldmath $\beta$}}_a^*} = \frac{ 
{\mbox{\boldmath $\beta$}}_a}{{{\mbox{\boldmath $\beta$}}_a^2}}  .
\end{equation} 
are the duals of the simple roots and the integers $n_a$ are the
topologically conserved charges corresponding to the homotopy class of
the scalar field at spatial infinity. For the BPS monopole
configurations, the magnetic charge is given so that $n_a \ge 0$.

Monopole solutions carrying a single unit of topological charge can be
obtained by simple embeddings of the $SU(2)$ solution. Each simple
root ${\mbox{\boldmath $\beta$}}_a$ defines an $SU(2)$ subgroup with
generators
\begin{eqnarray} 
t^1 &=& \frac{1 }{ \sqrt{2{{\mbox{\boldmath $\beta$}}_a^2}}} 
        \left(E_{{\mbox{\boldmath $\beta$}}_a} + 
        E_{-{\mbox{\boldmath $\beta$}}_a} \right), \nonumber \\
t^2 &=& -\frac{i}{ \sqrt{2{{\mbox{\boldmath $\beta$}}_a^2}}} 
        \left(E_{{\mbox{\boldmath $\beta$}}_a} - 
        E_{-{\mbox{\boldmath $\beta$}}_a} \right), \nonumber \\
t^3 &=&  {{\mbox{\boldmath $\beta$}}_a^*}  \cdot  {\bf H} . 
\label{sub}
\end{eqnarray}
If $\Phi^s({\bf r};v)$ and $A_i^s({\bf r};v)$ ($s=1,2,3$) is
the $SU(2)$ solution corresponding to a Higgs expectation value $v$,
then 
\begin{eqnarray} 
A_i({\bf r}) &=& \sum_{s=1}^3 A_i^s({\bf r};\,{\bf h}
\cdot {\mbox{\boldmath $\beta$}}_a ) t^s ,   \nonumber\\
\Phi({\bf r})&=& \sum_{s=1}^3 \Phi^s({\bf r};\,{\bf h}
\cdot {\mbox{\boldmath $\beta$}}_a ) t^s 
 + ( {\bf h} - {\bf h}\cdot {{\mbox{\boldmath $\beta$}}_a^*} \,\,
{\mbox{\boldmath $\beta$}}_a)\cdot {\bf H} , \label{embed}
\end{eqnarray}
is a solution with topological charges
\begin{equation}    
n_b =\delta_{ab}  ,
\end{equation}
and mass
\begin{equation}   
m_a = g {\bf h}\cdot {{\mbox{\boldmath $\beta$}}_a^*} .
\end{equation}
We will refer to such solutions as {\it fundamental monopoles} \cite{weinberg1}.
As in the $SU(2)$ case, there are dyon solutions corresponding to these
fundamental monopoles such that 
asymptotically
\begin{equation} 
{\bf E}^{(a)} =  \frac{q_a{\hat r}}{ 4\pi r^2 } 
 {{\mbox{\boldmath $\beta$}}_a^*}\cdot {\bf H }
\label{elecf}
\end{equation}
After quantization, the  electric charge $q_a$ is an integer multiple
of the unit $e|{{\mbox{\boldmath $\beta$}}_a^*}|^2$. {}From
Eqs.~(\ref{dyonsolution}) and (\ref{embed}), we get
the asymptotic  form of the Higgs field for such a dyon as $\Phi =
\Phi_0 +\Phi^{(a)} $ where 
\begin{equation} 
\Phi^{(a)}  =  - \frac{ {{\mbox{\boldmath
$\beta^*$}}_a}\cdot {\bf H } }{4\pi r}\sqrt{g^2+q_a^2}
\label{scalarf}
\end{equation}

There are a few difference between the $SU(2)$ and general cases in
the several BPS monopole solutions.  A notable difference, which will
be of importance later, concerns the symmetry of the two-monopole
solutions.  In the $SU(2)$ case these failed to be axially symmetric
because any gauge transformation that compensated the effects of a
spatial rotation on one of the monopoles shifted the phase of the
other monopole in the wrong direction.  By contrast, if the two are
different fundamental monopoles, it is always possible to find two
gauge transformations such that each gives the required phase shift on
one of the monopoles and leaves the other invariant.  As a result, the
superposition construction yields solutions with exact axial symmetry
\cite{ath}. This leads to an additional $U(1)$ symmetry in this case.

    A second difference from the $SU(2)$ case is the existence of
relatively simple, spherically symmetric solutions that can be
interpreted as superpositions of several fundamental monopoles at the
same point.  These are given by Eqs.~(\ref{sub}) and (\ref{embed}), 
but with a composite root ${\mbox{\boldmath $\alpha$}}$, rather than a 
simple root ${\mbox{\boldmath $\beta$}}_a$, defining the
$SU(2)$ subgroup.  The coefficients $n_a$ in the expansion
\begin{equation}   
    {\mbox{\boldmath $\alpha$}}^* =  \sum_{a=1}^k  
n_a {{\mbox{\boldmath $\beta$}}_a^*}  \label{composite-root}
\end{equation}
are the topological charges of the solution, while the mass is
\begin{equation} 
   m = \sum_a n_a m_a .
\end{equation}
Although the mass and topological charge of these solutions are
consistent with their interpretation as superpositions of several
noninteracting monopoles, one might still ask why these spherically
symmetric solutions should be viewed on such a different basis than
those obtained from the simple roots.  An answer is obtained by
counting the normalizable zero modes about these solutions.  After
gauge fixing, the number of zero modes about an arbitrary BPS solution
of magnetic charge (\ref{mcharge}) is \cite{weinberg1}
\begin{equation} 
      N = 4 \sum_{a=1}^k n_a .
\end{equation}
Thus, each of the fundamental monopoles has four zero modes, three
corresponding to spatial translations and the fourth to a global
$U(1)$ gauge rotation.  By contrast, the solutions based on composite
roots all have additional zero modes, with the number precisely that
expected if these are in fact superpositions of several fundamental
monopoles.

In the MSB phase there are as many charged vector bosons as the number
of positive roots, whose number is much larger than that of simple
roots if the gauge group is bigger than $SU(2)$.  On the other hand we
just argued that classically the number of fundamental monopoles is
identical to that of simple roots. This seems to contradict with the
duality hypothesis that the spectrum of charged particles should match
exactly that of magnetic monopoles.  However the duality is an
intrinsically quantum mechanical statement: one has to study quantum
mechanical spectrum of magnetic monopoles, which raises a possible
existence of the quantum mechanical bound state of fundamental
monopoles for each composite root. As the BPS mass formula is expected
to be exact in the $N=4$ supersymmetric theory, the bound energy of
these composite states would be zero.  This compels us to understand
the low energy dynamics of monopoles better.

The BPS configurations for a given magnetic charge, and so the same
energy, are parameterized by the $N$ collective coordinates $z_\alpha,
\alpha=1...N$, or moduli up to local gauge transformations.  For a
given magnetic charge, the space of gauge-inequivalent BPS
configurations is called the moduli space. A typical configuration is 
\begin{equation}
A_\mu({\bf x};z_\alpha) = ({\bf A}({\bf x};z_\alpha),
\Phi({\bf x},z_\alpha))
\end{equation}
with $A_4=\Phi$. The $4N$ zero modes $\delta_\alpha A_\mu = \partial
A_\mu /\partial z_\alpha - D_\mu \epsilon_\alpha $ satisfy the
linearized BPS equation and are chosen to satisfy the background gauge
$D_\mu \delta_\alpha A_\mu=0 $ with $\partial_4=0$.

When we consider fluctuations around the BPS magnetic monopole
configurations, there are massless modes and massive modes. If the
initial energy is arbitrary small, the dynamics of BPS monopoles may
be approximated by that of moduli\cite{manton}. The initial field
configuration at a given time will be characterized by $A_\mu({\bf
x},z_\alpha(t))$ and its time derivative, $\dot{z}_\alpha
\delta_\alpha A_\mu$ in the $A_0=0$ gauge. The Gauss law constraint on
the initial configuration is exactly the background gauge.

Since there is no force between monopoles at rest, one expects the low
energy dynamics is given by the kinetic part of the Yang-Mills-Higgs
Lagrangian. In the $A_0=0$ gauge, this becomes
\begin{equation}
{\cal L} = \frac{1}{2} G_{\alpha\beta}(z_\alpha)  \dot{z}_\alpha
\dot{z}_\beta
\label{modu}
\end{equation}
where  $G_{\alpha\beta}(z_\alpha) =\int d^3x \,{\rm tr}\,
\delta_\alpha A_\mu \delta_\beta A_\mu $. While one
can study some characteristics of this metric, it is hard to obtain
directly from the BPS field configurations which themselves are not
known in general.  However some formal characteristics of the metric
can be deduced from this.  The important property of the metric is
that it is hyper-K\"ahler. When we consider the supersymmetric
Yang-Mills theory, the original supersymmetry which should be
incorporated by supersymmetrizing the above Lagrangian \cite{blum}.

A $4n$-dimensional manifold with a metric is hyper-K\"ahler if it
possesses three covariantly constant complex structures ${\cal
J}^{(k)}, k=1,2,3$ that also form a quarternionic structure and if
the metric is pointwise Hermitian with respect to each ${\cal
J}^{(k)}$. If we recast the zero modes $A_\mu$
into a spinor $\Psi=\delta
A_4+i\tau_i\delta A_i$ where $\tau_i$'s are the Pauli matrices, the
zero mode equation is manifestly invariant under right multiplications
by the $i\tau_k$'s \cite{weinberg1}, and this induces the almost
quarternionic structure on the moduli space. Detailed arguments that
the moduli space is hyper-K\"ahler can be found in
Refs.~\cite{atiyah,jg}.

\section{Distinct Fundamental Monopoles}

The metric on the moduli space determines the motions of slowly moving
monopoles.  Conversely, the form of the moduli space metric can be
inferred from a knowledge the interactions between monopoles.  These
become quite complicated when several monopoles approach one another
as their nonabelian cores overlap.  However, for finding the metric in
the regions of the moduli space corresponding to large intermonopole
distances, it is sufficient to examine the long-range pairwise
interactions between widely separated monopoles. In large separation
of monopoles, the electric charge of individual monopole, the momentum
for the phase zero mode, is conserved and the interaction between $n$
monopoles of $4n$ coordinates becomes that between dyons of $3n$
coordinates. This analysis has been carried out previously for $SU(2)$
\cite{gibbons1} and generalized to larger gauge groups in
Ref.\cite{klee1}.

We begin by considering the interactions between a single pair of
fundamental dyons, with positions ${\bf x}_a$ and velocities ${\bf
v}_a$ and carrying magnetic charges $g {{\mbox{\boldmath
$\beta$}}_a^*} $ and electric charges $q_a {{\mbox{\boldmath
$\beta$}}_a^*}$ ($a = 1 ,2$).  If the separation between the dyons is
much larger than the radius of a monopole core, the electromagnetic
interactions between them can be well approximated by the standard
results for a pair of moving point charges.  There is also a
long-range scalar force that is manifested as a position-dependent
shift in the dyon mass.  Recall that the mass of an isolated dyon is
$\tilde{m}_a = {\bf h} \cdot {{\mbox{\boldmath $\beta$}}_a^*}
\sqrt{g^2 + q_a^2}$.  When there is a second dyon present, its
correction to the scalar field must be added to ${\bf h}$ in this
formula.

The electric and scalar fields for dyon 2 at the origin are given by
Eqs.~(\ref{elecf}) and (\ref{scalarf}). The magnetic field for dyon 2 at
the origin is ${\bf B}^{(a)}$ obtained from Eq.(\ref{magf}) with only
$n_a=1$ nonvanishing.  The four potential ${\bf A}^{(2)}$ and
$A_0^{(2)}$ for dyon 2 can be obtained from the electromagnetic field.
The dual four potential $\tilde{\bf A}^{(2)}$ and $\tilde A_0^{(2)}$
is defined so that ${\bf E} = - {\bf \nabla} \times \tilde {\bf A}$ and
${\bf B} = -{\bf \nabla} \tilde A_0 - \partial \tilde{\bf A}/\partial
t$.  The scalar field and  four-vector  potentials can be generalized to
the Lienard-Wiechart forms by taking into account the motion of dyon
2\, \cite{gibbons1}.

The motion of dyon 1 in the background of the fields generated by dyon
2 is described by the Lagrangian 
\begin{eqnarray}
L^{(1)} &=& \sqrt{g^2 + q_1^2} \,\, {\rm tr} \left\{ 
{{\mbox{\boldmath $\beta$}}_1^*} \cdot {\bf H} \bigl[ \Phi_0 +
\Phi^{(2)}({\bf x}_1)\bigr] \right\} \sqrt{1- {\bf v}_1^2}\nonumber \\
&+& q_1 \,\,
{\rm tr} \left\{ {{\mbox{\boldmath $\beta$}}_1^*} \cdot {\bf H} \bigl[
{\bf v}_1 \cdot {\bf A}^{(2)}({\bf x}_1) - A_0^{(2)}({\bf x}_1)
\bigr]\right\} \nonumber \\ &+& g \,\, {\rm tr} \left\{
{{\mbox{\boldmath $\beta$}}_1^*} \cdot {\bf H}\bigl[ {\bf v}_1 \cdot
\tilde{\bf A}^{(2)}({\bf x}_1)  -  \tilde A_0^{(2)}({\bf x}_1)
\bigr] \right\}. \label{one} 
\end{eqnarray}
Substituting the Lienard-Wiechart form of the potentials and scalar
field into Eq.~(\ref{one}) and keeping only terms of up to second
order in $q_j$ or ${\bf v}_j$, we obtain
\begin{eqnarray}
L^{(1)}  &=& -m_1  \left( 1 - \frac{1}{2}{\bf v}_1^2 + 
\frac{q_1^2 }{2 g^2} \right) \nonumber\\
& -& \frac{g^2 }{ 8\pi r_{12} } {\mbox{\boldmath $\beta$}}^*_1 \cdot
{\mbox{\boldmath $\beta$}}^*_2   \left[({\bf v}_1 -{\bf v}_2)^2
- \frac{(q_1-q_2)^2}{g^2} \right] \nonumber \\
&-&\frac{g}{ 4\pi } {\mbox{\boldmath $\beta$}}^*_1 \cdot
{\mbox{\boldmath $\beta$}}^*_2   (q_1 - q_2) ({\bf v}_1 -{\bf
v}_2)\cdot {\bf w}_{12}, 
\end{eqnarray}
where $m_1 = g{\bf h}\cdot {\mbox{\boldmath $\beta$}}^*_1$,  $r_{12}=
|{\bf x}_1 - {\bf x}_2|$ and ${\bf 
w}_{12} \equiv{\bf w}({\bf x}_1 -{\bf x}_2) $ is a Dirac monopole
potential, defined so that   
\begin{equation}
    {\bf \nabla}  \times {\bf w}({\bf x})  = - \frac{ {\bf x} }{ |{\bf
x}|^3} . 
\end{equation}
The Lagrangian describing the dynamics of slowly moving two dyons at
large separation follows if one symmetrizes  the above Lagrangian
by adding  the noninteracting part for dyon 1.

The extension to an arbitrary number of fundamental dyons is
straightforward, provided that their mutual separations are all large.
The Lagrangian obtained by adding all the pairwise interactions can be
written as
\begin{equation}
L = \frac{1}{2} M_{ab} \left( {\bf v}_a \cdot {\bf v}_b 
      - {q_a q_b \over g^2}   \right)  + \frac{g }{4\pi} q_a
      {\bf W}_{ab}\cdot {\bf v}_b ,
\end{equation}
where
\begin{equation}
M_{aa} = m_a  - \sum_{c\ne a} \frac{g^2 {\mbox{\boldmath $\beta$}}_a^* 
\cdot {\mbox{\boldmath $\beta$}}_c^*}{ 4\pi r_{ac}},\,\,\,\,\,
M_{ab} =\frac{g^2 {\mbox{\boldmath $\beta$}}_a^* \cdot 
{\mbox{\boldmath $\beta$}}_b^*}{ 4\pi r_{ab}}\quad
\hbox{\hskip 1cm if $a\neq b$},
\end{equation}
with $m_a=g\,\mbox{\boldmath $\beta$}_a^*\cdot {\bf h}$, and
\begin{equation}
{\bf W}_{aa}=-\sum_{c\neq a}{\mbox{\boldmath $\beta$}}_a^*\cdot
{\mbox{\boldmath $\beta$}}_c^*{\bf w}_{ac},\,\,\,\,
{\bf W}_{ab}={\mbox{\boldmath $\beta$}}_a^*\cdot
{\mbox{\boldmath $\beta$}}_b^*{\bf w}_{ab}\quad
\hbox{\hskip 1cm if $a\neq b$}.
\end{equation}
with ${\bf w}_{ab}$ being value at ${\bf x}_a$ of the Dirac potential
due to the $b$th monopole. The  monopole
rest masses have been omitted.

     To obtain the moduli space metric, we need a Lagrangian that is
purely kinetic; i.e., one in which all terms are quadratic in velocities. 
This can be done by interpreting the $q_b/e$ as conserved momenta conjugate
to cyclic angular variables $\xi_b$. Because  
$q_b/e$ is quantized in  integer multiples of ${\mbox{\boldmath
$\beta$}}_b^2$, the period of $\xi_j$ must be $2\pi/{\mbox{\boldmath
$\beta$}}_b^2$. Thus, if we make the identification
\begin{equation}
q_a/e = \frac{g^4}{16\pi^2}(M^{-1})_{ab}(\dot\xi_b + 
{\bf W}_{bc}\cdot {\bf v}_c ),
\end{equation}
the desired Lagrangian ${\cal L}$ is the Legendre
transform  
\begin{eqnarray}
{\cal L}&=& L + \sum_j \dot{\xi}_bq_b/e \nonumber \\
 &=& \frac{1}{2} M_{ab}  {\bf v}_a \cdot {\bf v}_b +  
    \frac{g^4}{2(4\pi)^2}(M^{-1})_{ab} 
    \left( \dot\xi_a + {\bf W}_{ac}\cdot {\bf v}_c \right)
    \left( \dot\xi_b + {\bf W}_{bd}\cdot {\bf v}_d \right). \label{L}
\end{eqnarray} 
{}From this we immediately obtain the large separation 
approximation to the moduli space metric,
\begin{equation}
{\cal G}=M_{ab}d{\bf x}_a\cdot d{\bf x}_b+\frac{g^4}{16\pi^2} 
(M^{-1})_{ab}(d\xi_a+{\bf W}_{ac}\cdot d{\bf x}_c)(d\xi_b+{\bf W}_{bd}
\cdot d{\bf x}_d) .\label{metric}
\end{equation}
Note that this metric is equipped with a number of $U(1)$ isometries, each
of which is generated by the constant shift of one of the $\xi_b$'s.

The fact that this asymptotic metric is hyper-K\"ahler can be shown
trivially, following the argument by Gibbons and
Manton\cite{gibbons1}. The key ingredient is that $\nabla 1/r = \nabla
\times {\bf w}(r)$. The question is whether the asymptotic metric when
it is extended in the interior region is nonsingular.

For the case of $SU(2)$, one can easily see that this approximation 
to the moduli space metric cannot be exact \cite{gibbons1}.  First of
all, it develops singularities if any of the intermonopole distances
becomes too small, whereas the moduli space metric should be
nonsingular.  Second, for the case of two monopoles the approximate
metric is independent of the relative phase angle $\xi_1
-\xi_2$.  If this isometry were exact, the two-monopole solutions
would be axially symmetric, which we know is not the case.  

Neither of these objections arises for the moduli space corresponding
to a collection of several distinct fundamental monopoles in a larger
group, provided that each corresponds to a different simple root.  The
metric can be simplified by going to the center of mass frame for 
$r$ fundamental monpoles of mass $m_a$ for
connected simple roots ${\mbox{\boldmath $\beta$}}_a$ of a simple
group $G$. (For a given number of the distinct monopoles, we choose
for the convenience the minumum size gauge group which allows them.)
The center-of-mass position and total charge are given as
\begin{eqnarray}
 {\bf R} &=& \frac{1}{M} \sum_a m_a {\bf x}_a \\
  q_\chi &=&  \frac{1}{eM} \sum_a m_a q_a 
\end{eqnarray}
with the total mass $M=\sum_a m_a$.  The conjugate global phase of the
total charge $q_\chi$ is $\chi = \sum_a \xi_a$ whose value lies on the
real line $R$.  We label by $A=1,,,r-1$ the $r-1$ links between the
adjacent pairs of simple roots, which can be read from the Dynkin
diagram. The relative coordinates and charges between monopoles are
defined as
\begin{eqnarray}
{\bf r}_A &=& {\bf x}_a-{\bf x}_b \\
q_A  &=& \frac{\lambda_A}{2e} (q_a-q_b)
\end{eqnarray}
where ${\mbox{\boldmath $\beta$}}_a$ and  ${\mbox{\boldmath
$\beta$}}_b$ are connected by the rink $A$ and $\lambda_A =
-2{\mbox{\boldmath $\beta$}}_a^*\cdot{\mbox{\boldmath $\beta$}}_b^*$.
The conjugate angular variable $\psi_A$ for each $q_A$ has the range
$[0,4\pi]$.

The center-of-mass part of the moduli space metric (\ref{metric})
is the
\begin{equation}
{\cal G}_{\rm cm}=M \left(d{{\bf R}}^2 + \frac{g^4
}{16\pi^2 M^2 }d{\chi}^2\right),
\label{com}
\end{equation}
and is the metric of a flat four-dimensional manifold.  The metric
\cite{klee2} for the relative coordinates obtained from Eq.
(\ref{metric}) is
\begin{equation}
{\cal G}_{\rm rel} = C_{AB}d{\bf r}_A\cdot d{\bf r}_B 
 + \frac{g^4\lambda_A\lambda_B}{64\pi^2}(C^{-1})_{AB}D\psi_A D\psi_B
\end{equation}
where $D \psi_A = d\psi_A + {\bf w} ({\bf r}_A)\cdot d {\bf r}_A$
and the $(r-1)\times (r-1)$ matrix $C_{AB}$ is
\begin{equation}
C_{AB} = \mu_{AB} + \delta_{AB}\frac{g^2\lambda_A}{8\pi r_A}
\end{equation}
with $r_A=|{\bf r}_A|$ and the reduced mass matrix $\mu_{AB}$ can be
deduced from the free Lagrangian. The total moduli space is not a
simple direct product of the center-of-mass part $R^4$ and the
relative part.  {}From the translations generated from $\xi_i
\rightarrow \xi_i + 4\pi$, the space $R\times S_1^{r-1}$ of the
$(\chi,\psi_A)$ should be divided by an indentification map of integer
group $Z$.

The metric is rotationally invariant and so there is a conserved
angular momentum in the dynamics, which can also be written as a sum
of ${\bf J}_{\rm cm}=M{\bf R}\times \dot{\bf R}$ and
\begin{equation}
{\bf J}_{\rm rel}  = \sum_{AB} C_{AB} {\bf r}_A\times \dot{\bf r}_B +
\sum_A q_A \hat{\bf r}_A
\end{equation}

While our metric is obtained by considering the monopoles in large
separations, this metric ${\cal G}_{\rm rel}$ is smooth everywhere,
complete and has the right symmetry. It is not only invariant under
the spatial rotation but also under the $r-1$ $U(1)$ phase shifts
$\psi_A\rightarrow \psi_A + {\rm constant}$.  The origin where $r_A=0$
is the point where all monopoles come together, which should be
spherically symmetric as it can be representated by the $SU(2)$
embedding solution mentioned earlier with the maximum positive root
$\sum_i {\mbox{\boldmath $\beta$}}$. This leads one to suspect our
metric is the right metric everywhere. This conjecture has been argued
to be correct lately  by Murray and Chalmers \cite{murray,chalmers}.

In particular when $r=2$, we can confirm this explicitly. The relative
moduli space is a four-dimensional hyper-K\"ahler space with
rotational symmetry.  Atiyah-Hitchin\cite{atiyah} showed that such a
space is one of the following manifolds:

\vskip 2mm
\noindent
1) flat $R^4$.

\noindent
2) the Taub-NUT geometry with an $SU(2)$ rotational
isometry.  

\noindent
3)  the Atiyah-Hitchin geometry with an $SO(3)$
rotational isometry.

\noindent
4) the Eguchi-Hanson gravitational instanton \cite{eguchi}.

\vskip 2mm

The only one with the right symmetry and interaction, the asymptotic
form and the symmetric point turns out to be the Taub-NUT metric
\cite{klee2,gaunt,connell}.

\section{Discussion}

In this talk we have reviewed the method to derive the moduli space
metric which describes the low energy dynamics of the $n$ distinct
fundamental BPS monopoles in the Yang-Mills-Higgs systems with
arbitrary gauge group. The asymptotic metric obtained by studying the
low energy dynamics of dyons in large separation seems to be correct
everywhere as it has the right isometries and properties, like
hyper-K\"ahlerness, smoothness, and completeness. We then focused
ourselves to the moduli space metric for two monopoles, whose
correctness everywhere can be seen easily.

There are several questions raised from this point. One interesting
issue concerns theories for which there is the nonabelian unbroken
gauge group.  There has been a considerable progress along this
direction by us \cite{klee3}, which is summerized in Erick's talk.  As
discussed in detail there, the moduli space of monopoles with total
magnetic charge purely abelian in this case is the appropriate limit
of the moduli space of the identical magnetic charge in the MSB case.
The massless monopoles does not correspond to distinct solitons but
instead manifest themselves as {\it nonabelian cloud} surrounding the
massive monopoles.  While the zero modes for massive monopoles remain
to be interpreted as the positions and abelian phases, the zero modes
corresponding to the massless monopoles are transformed into an equal
number of nonabelian global gauge orientation and gauge-invariant
structure parameters characterizing the nonabelian cloud. The unbroken
nonabelian gauge group becomes the part of the isometry group of the
moduli space.

The second question is whether one can extend the above work to the
many-monopole case where some of monopoles are identical.  The
asymptotic form for the metric develops a curvature singularity if a
pair of identical monopoles are brought too close together.  Recently
there seems to be some progress for the case where the number of
identical monopoles are less than or equal to two by using the
Legendre transformation \cite{chalmers}. If this is successful, one
further expects some simplifications in the cases involving two or more
identical monopoles where exist of spherically symmetric solutions.
Further progress along this direction would be quite interesting
as it may shed some light on how nonabelian cloud structures itself
around the sources.

Finally, one of the more compelling motivations for studying the low
energy dynamics of monopoles and dyons arises from the Montonen-Olive
duality conjecture \cite{montonen}.  In certain supersymmetric Yang-Mills
theories, duality maps strongly coupled electric theories to weakly
coupled magnetic ones, thus enabling one to probe the nonperturbative
nature of strongly coupled Yang-Mills theories.  A notable example of
this is the $N=2$ supersymmetric gauge theories softly broken to
$N=1$, where confinement is explicitly realized through magnetic
monopole condensation \cite{seiberg}.

In order for such a duality to make sense, however, the spectrum of
magnetically charged particles must be consistent with that predicted
by the duality mapping of the electrically charged ones.  In $N=4$
supersymmetric Yang-Mills theories, duality relates each elementary
massive vector meson of electric charge $\mbox{\boldmath $\gamma$}$ to
a tower of dyons of magnetic charge $\mbox{\boldmath $\gamma$}^*$
among others, where $\mbox{\boldmath $\gamma$}$ is any root of the
gauge algebra.  The analysis of the classical solutions tells us,
however, that a monopole based on a composite root is not a
fundamental entity, but rather corresponds to a mere coincidence point
on a multi-monopole moduli space. Thus, it is imperative to see if the
quantization of the multi-monopole dynamics leads to a bound state
that could conceivably be dual to the vector meson whenever
$\mbox{\boldmath $\gamma$}$ is composite.

This program has recently been carried out in the MSB cases. The
bound state of two distinct monopoles in the MSB case was found by the
authors \cite{klee2}, and also independently by Gauntlett and Lowe
\cite{gaunt}. The most massive of the charged vector mesons in this
theory is dual to a threshold bound state with composite magnetic
charge, which is realized as a unique normalizable harmonic form on
the Taub-NUT manifold.  The threshold bound states of any number of
distinct monopoles have been found recently by Gibbons
\cite{gibbons2}.  In theories with unbroken nonabelian gauge symmetry,
the duality implies the existence of many threshold bound states whose
magnetic charges are be abelian or nonabelian~\cite{klee3}. Since some
candidate of the moduli space metric is known explicitly, it would be a
challenge to find such bound states in this case.

\vskip 1cm
\centerline{\large\bf Acknowledgments}
\vskip  5mm

We thank the organizers, especially Peter Tinyakov and Valery Rubakov,
of the Quark-96 conference at Yaroslavl, Russia for warm hospitality
and pleasant atmosphere.  This work was supported in part by the NSF
Presidential Young Investigator program.


\begin{thebibliography}{99}


\bibitem{bps}{E.B. Bogomol'nyi, Sov. J. Nucl. Phys. {\bf
24}, 449 (1976); M.K. Prasad and C.M. Sommerfield, Phys. Rev. Lett.
{\bf 35}, 760 (1975); S. Coleman, S. Parke, A. Neveu and C.M.
Sommerfield, Phys. Rev. {\bf D15}, 544 (1977).}

\bibitem{manton}{N.S. Manton, Phys. Lett. {\bf 110B}, 54 (1982).}


\bibitem{montonen}{C. Montonen and D. Olive, Phys. Lett. {\bf 72B}, 117
(1977); H. Osborn, Phys. Lett. {\bf 83B}, 321 (1979).} 


\bibitem{olive}{D. Olive, {\it Exact Electromagnetic Duality},
hep-th/9508089; J.A. Harvey, {\it Magnetic Monopoles, Duality and
Supersymmetry}, hep-th/9603986}


\bibitem{atiyah}{M.F. Atiyah and N.J. Hitchin, {\it The Geometry and 
Dynamics of Magnetic Monopoles}, Princeton Univ. Press, Princeton (1988);
Phys. Lett. {\bf 107A}, 21 (1985); Phil. Trans. R. Soc. Lon. {\bf
A315}, 459 (1985).}

\bibitem{sen}{A. Sen, Phys. Lett.  {\bf B329}, 217 (1994); S. Sethi, M.
Stern and E. Zaslow,   Nucl. Phys. {\bf B457}, 484 (1995); J.P.
Gauntlett and J. Harvey, Nucl. Phys. {\bf B463},  287 (1996).} 

\bibitem{porrati}{M. Porrati, Phys. Lett. {\bf B377}, 67 (1996).}


\bibitem{gibbons1}{ N.S. Manton, Phys. Lett. {\bf B 154}, 397 (1985);
G.W. Gibbons and N.S. Manton, Phys. Lett. {\bf B356}, 32 (1995).}


\bibitem{weinberg1}{E.J. Weinberg, Nucl. Phys.  {\bf B167}, 500 (1980).}


\bibitem{klee1}{K. Lee, E.J. Weinberg and P. Yi, Phys. Rev. {\bf D54}, 1633 
(1996).}

\bibitem{klee2}{K. Lee, E.J. Weinberg and P. Yi, Phys. Lett. {\bf B376}, 97
(1996).} 

\bibitem{gaunt}{J.P. Gauntlett and D.A. Lowe, Nucl. Phys. {\bf B 472},
194 (1996).}

\bibitem{connell}{S.A. Connell, {\it
The dynamics of the $SU(3)$ charge (1,1) magnetic monopoles,} University
of South Australia preprint.}



\bibitem{murray}{M.K. Murray, {\it A note on the $(1,1,\dots,1)$ monopole
metric}, hepth/9505054, a University of Adelaide preprint.}

\bibitem{chalmers}{G. Chalmers, {\it Multi-monopole moduli spaces for $SU(N)$
gauge group}, ITP-SB-96-12, hep-th/9605182.}


\bibitem{guth}{A.H. Guth and E.J. Weinberg, Phys. Rev. {\bf D14}, 1660
(1976).}

\bibitem{klee3}{K. Lee, E.J. Weinberg and P. Yi, {\it Massive and
Massless Monopoles with Nonabelian Magnetic Charges}, CU-TP-55 and
hep-th/9605229.} 



\bibitem{topology}{P. Goddard,  J. Nuyts and D. Olive, Nucl. Phys. {\bf
B125}, 1 (1977); F. Engert and P. Windey, Phys. Rev. {\bf D14}, 2728 (1976).}

\bibitem{ath}{C. Athorne, Commun. Math. Phys. {\bf 88}, 43 (1983).}


\bibitem{blum}{J.D. Blum, Phys. Lett.  {\bf B333} (1994) 92; J.
Gauntlett, Nucl. Phys. {\bf B411}, (1994) 443.}  

\bibitem{jg}{J. Gauntlett, Nucl. Phys. {\bf B411}, 443 (1994).} 



\bibitem{gibbons2}{G.W. Gibbons, Phys. Lett. {\bf B282}, 53 (1996).}


\bibitem{eguchi}{ T. Eguchi and A.J. Hanson, Ann. Phys. {\bf 120}, 82
(1979).} 




\bibitem{seiberg}{N. Seiberg and E. Witten, Nucl. Phys. {\bf B426}, 19
(1994); (E) {\bf B430}, 485 (1994); {\it ibid}. {\bf B431}, 484 (1994).}





\end{thebibliography}
\end{document}